\documentclass{article}

\input{tcilatex}

\begin{document}

\textbf{DISTRIBUTED SOURCES, ACCELERATED UNIVERSE, CONSCIOUSNESS AND QUANTUM
ENTANGLEMENT}

\bigskip

E. A. Novikov

\bigskip

Institute for Nonlinear Science, University of California - San Diego, La
Jolla, CA 92093 - 0402

\bigskip

It is shown that such diverse phenomena as accelerated Universe,
consciousness and quantum entanglement are connected by the concept of
distributed sources and imaginary (particularly, tachyon) fields.

\bigskip

The singular vortices, sources (sinks) and vortex-sinks are well known
models in fluid dynamics (including geophysical fluid dynamics), magnetized
plasma, superfluidity and superconductivity (see, for example, Refs. 1 - 3
and references therein). Dynamics of distributed vortices is a well
developed area of research. However, to our knowledge, dynamics of
distributed sources (DS) has not been considered until recently [4]. One of
the applications of DS, indicated in Ref. 4, is cosmology. Particularly,
solution of corresponding equation with constant intensity of DS [4] is
similar to homogeneous solution of general relativity with the cosmological
constant (CC). However, analysis of DS in Ref. 4 was nonrelativistic. In
this Note we consider relativistic generalization of DS. We will show that
DS, in certain simple case, produce the effect of CC in the accelerated
expansion of the Universe. Relativistic DS can be used also for local
phenomena in cosmology. We will also show a connection between DS and
phenomena of consciousness and quantum entanglement.

\bigskip Local intensity of nonrelativistic DS is characterized by the
divergency of the velocity field $\ \partial v^{\alpha }/\partial x^{\alpha
} $ ( summation over the repeated Greek indexes is assumed from 1 to 3 ).
Dynamical equation for DS was obtained by considering superposition of
localized sources, which move each other with induced velocity field [2 - 4].

For relativistic DS we need four-dimensional velocity field (see, for
example, Refs. 5, 6):

\begin{equation}
u^{i}=\frac{dx^{i}}{ds}=\gamma \frac{dx^{i}}{d\tau },\;\gamma
=(1-v^{2})^{-1/2},\;\frac{d}{ds}\equiv u^{k}\frac{\partial }{\partial x^{k}}
\tag{1}
\end{equation}%
Here position of fluid element is characterized by the 4-vector $x^{i}$ with
components $(\tau ,x^{\alpha })$, where $\tau =ct$ and $c$ is the velocity
of light. Components $x^{\alpha }$ in turn can be considered as functions of 
$\tau $ and some identification parameters, for example, initial positions $%
x_{o}^{\alpha }$. 4-vector $u^{i}$ has components $(\gamma ,\gamma v^{\alpha
})$ with $v^{\alpha }$ normalized by $c$, $\gamma $ is the Lorentz factor
and summation over repeated Latin indexes is assumed from 1 to 4. Components 
$u^{i}$ can be considered as functions of $x^{i}$ or as functions of $(\tau
,x_{o}^{\alpha })$. For simplicity, we will use the covariant
differentiation only where it is needed (see below equation (8)).

It seems natural to characterize local intensity of relativistic DS by
divergency of 4-velocity:

\begin{equation}
\sigma =\frac{\partial u^{i}}{\partial x^{i}}=\frac{\partial \gamma }{%
\partial \tau }+\frac{\partial (\gamma v^{\alpha })}{\partial x^{\alpha }} 
\tag{2}
\end{equation}

In nonrelativistic case with $v\ll 1$, we have: $\gamma \approx 1,\;\partial
\gamma /\partial \tau \approx v(\partial v/\partial \tau )$; the first term
in the right hand side of (2) is much smaller than the second term and we
return to the presented above nonrelativistic expression (apart from
normalization by $c$).

The equation for $\sigma $, which is relativistic generalization of
corresponding equation in Ref. 4, reads:

\begin{equation}
\frac{d\sigma }{ds}=f  \tag{3}
\end{equation}

Here $f$ may include diffusion and some other effects (see Ref. 4). In this
Note we consider "free" sources with $f=0$ (see below a justification of
such choice for a concrete physical problem). In this case, taking into
account (2), we get equation:

\begin{equation}
\frac{\partial \gamma }{\partial \tau }+\frac{\partial (\gamma v^{\alpha })}{%
\partial x^{\alpha }}=\sigma _{o}  \tag{4}
\end{equation}

Here $\sigma _{o}$ , generally, depends on $x_{o}^{\alpha }$. In Ref. 4 some
analytical solutions of nonrelativistic analog of equation (4) were obtained
for certain initial conditions, characterized by field $\sigma _{o}$.
Corresponding analytical solutions of equation (4) can also be obtained. In
this Note we consider one important and specifically relativistic case.

What we have in mind is the problem of accelerated Universe (AU). The
accelerated expansion is extracted from observed luminosity of the type Ia
supernovae [7 - 9]. The described above DS model suggests production of
particles in AU. Balance of the proper number density of particles $n$ can
be written in the form:

\begin{equation}
\frac{\partial (u^{i}n)}{\partial x^{i}}=\frac{dn}{ds}+\sigma n=q  \tag{5}
\end{equation}%
where $q$ is the rate of particle production. In a quasistationary case $%
\sigma n\approx q$ and the macroscopic fluid, consisting of particles, can
be considered approximately incompressible. However, this is not an ordinary
fluid, as was stressed in Ref. 4. The traditional consideration with the
energy-momentum tensor has to be modified. The system is not Hamiltonian, as
was indicated in the case of the system of interacting point sources [2]. It
seems natural to assume that in the quasistationary case $q$ is
approximately proportional to $n$, which means that $\sigma \approx \sigma
_{0}$. This is a justification of the choice \ $f=0$. More general cases can
be studied in future.

Let us consider homogeneous AU by using the thermodynamic relation:

\begin{equation}
dE=\delta Q-pdV  \tag{6}
\end{equation}%
where $E$, $p$, $V$, are the energy, pressure, and the volume of the system,
and $\delta Q$ is a supply of energy by production of particles. The
production of particles can be considered as some sort of internal boundary
condition. For homogeneous AU we can write:

\begin{equation}
\delta Q=\varepsilon _{s}dV  \tag{7}
\end{equation}%
where $\varepsilon _{s}$ is the energy density produced by DS. For dustlike
matter $\varepsilon _{s}=mnc^{2}$, where $m$ is the mass of corresponding
particles. Combining (6) and (7), we see that effect of DS in the described
case is equivalent to the negative pressure $-\varepsilon _{s}$. The
negative pressure, in turn, corresponds to the effect of CC (see, for
example, Ref. 9).

At the last stage of AU, when DS will dominate the dynamics, we can
determine the scale factor $a(\tau )$ from equation (4) written in the
covariant form:

\begin{equation}
u_{;\;i}^{i}=\frac{\partial u^{i}}{\partial x^{i}}+\Gamma
_{ki}^{i}u^{k}=\sigma _{o}  \tag{8}
\end{equation}

The Christoffel symbols satisfy condition [5]:

\begin{equation}
\Gamma _{ki}^{i}=\frac{1}{2g}\frac{\partial g}{\partial x^{k}}  \tag{9}
\end{equation}%
where $g$ is the determinant of the metric tensor. For homogeneous isotropic
AU, in the proper synchronized frame of reference [5] we have: $%
u^{o}=1,u^{\alpha }=0$. From (8) and (9) we get:%
\begin{equation}
\frac{\partial g}{\partial \tau }=2\sigma _{o}g  \tag{10}
\end{equation}%
Taking into account that $g\sim a^{6}$ and $\sigma _{o}$ is constant (for
isotropic AU), we have:

\begin{equation}
a(\tau )=a_{o}\exp \{\frac{1}{3}\sigma _{o}\tau \}  \tag{11}
\end{equation}%
This simple result conforms to the nonrelativistic consideration [4] and to
the analogy with CC.

Thus, DS in the simplest case reproduce the effect of CC. In addition, DS
can provide a novel intuition and a quantitative analysis not only for the
global description of AU, but also for more local (nonhomogeneous and
nonquasistationary) cosmological phenomena.

Asymptotically, expansion (11) becomes superluminary: $da/d\tau >1$. This
suggests that particles and fields created by DS have imaginary
(particularly, tachyon) components. These imaginary fields (IF) can play an
important role not only in cosmology, but also in a variety of phenomena.

Consider the renormalization procedures [10], which are the basis for
substantial part of contemporary physics. Feynman, who (with Tomonaga and
Schwinger) got the Nobel Prize for refinement and applications of these
procedures, compared renormalization with sweeping dust (infinities) under
the rug.

In recent note [11] IF have been used to eliminate divergencies in the
classical theory of electromagnetic field, namely, infinite self-energy of
electrons and paradoxical self-acceleration. The natural next step is to
eliminate infinities in quantum field theories with an appropriate use of IF.

Another application of IF is to the phenomena of consciousness (considered
as collective effect of billions of interconnected nonlinear neurons). The
brain activity revealed the regime of scale-similarity [12-14], which is
typical for systems with strong interaction of many degrees of freedom
(particularly, for turbulence [15]). Modeling of the effects of
consciousness on the electric currents in the human brain leads to the use
of IF [16]. Possible connection of DS with consciousness was indicated in
Ref. 16. Now we have additional reason (11) in favor of such connection.
Another possible connection is between IF, which eliminate electromagnetic
divergencies [11], and IF in the modeling of consciousness [16].

Finally, consider the quantum entanglement, the EPR experiment and all that
(see corresponding discussions in Refs.17, 18). It seems natural to assume
that relativistic DS are produced by the quantum vacuum. According to (11),
DS lead to superluminary IF-effect on cosmological scale. Why the same
quantum vacuum can not produce locally an IF-signal in special
circumstances, say, during the process of quantum measurement? This will
explain the quantum entanglement.

Perhaps, DS \& IF provide the missing link between the quantum theory and
the general relativity.

\bigskip

\bigskip \textbf{References}

\bigskip

[1] E. A. Novikov, Ann. N. Y. Acad. Sci. \textbf{357}, 47 (1980)

[2] E. A. Novikov and Yu. B. Sedov, Fluid Dyn. \textbf{18}, 6 (1983)

[3\} A. E. Novikov and E. A. Novikov, Phys. Rev. E \textbf{54}, 3681 (1996)

[4] E. A. Novikov, Phys. of Fluid \textbf{15} (9), L65 (2003);
arXiv:nlin.PS/0501010

[5] L. D. Landau and E. M. Lifshitz, The Classical Theory of Fields,
Pergamon Press, 1987

[6] L. D. Landau and E. M. Lifshitz, Fluid Mechanics, Butterworth-Heinemann,
1995

[7] A. G. Riess et al., Astron. J. \textbf{116}, 1009 (1998); Astrophys. J., 
\textbf{607}, 665 (2004)

[8] S. Perlmutter et al., Astrophys. J. \textbf{517}, 565 (1999)

[9] Reviews of theories of AU with many references can be found in V.
Faraoni, Cosmology in Scalar-Tensor Gravity, Kluwer, 2004, and in D. H.
Coule, Class. Quantum Grav. \textbf{22}, R125 (2005).

[10] V. B. Berestetskii, E. M. Lifshitz and L. P. Pitaevskii, Quantum
Electrodynamics, Pergamon Press, 1982

[11] E. A. Novikov, arXiv:nlin.PS/0509029

[12] E. Novikov, A. Novikov, D. Shannahof-Khalsa, B. Schwartz, and J.
Wright, Phys. Rev. E \textbf{56}(3), R2387 (1997)

[13] E. Novikov, A. Novikov, D. Shannahof-Khalsa, B. Schwartz, and J.
Wright, in Appl. Nonl. Dyn. \& Stoch. Systems (Ed. J. Kadtke \& A. Bulsara),
p. 299, Amer. Inst. Phys., N. Y., 1997

[14] E. S. Freeman, L. J. Rogers, M. D. Holms, D. L. Silbergelt, J.
Neurosci. Meth. \textbf{95}, 111 (2000)

[15] E. A. Novikov, Phys. Rev. E \textbf{50}(5), R3303 (1994)

[16] E. A. Novikov, arXiv:nlin.PS/0309043; arXiv:nlin.PS/0311047;
arXiv:nlin.PS/0403054; Chaos, Solitons \& Fractals, v. 25, p. 1-3 (2005);
arXiv:nlin.PS/0502028

[17] D. Bohm \& B. J. Hiley, The Undivided Universe, Routledge 1993

[18] Roger Penrose, The Road to Reality, Jonathan Cape 2004

\bigskip 

\end{document}